\documentstyle[prl,aps,epsf,floats]{revtex}
\tighten

\def\ltap{\ \raise.3ex\hbox{$<$\kern-.75em\lower1ex\hbox{$\sim$}}\ }
\def\gtap{\ \raise.3ex\hbox{$>$\kern-.75em\lower1ex\hbox{$\sim$}}\ }

\def\eV{{\rm eV}}

\def\GeV{{\rm GeV}}

\def\PLB#1#2#3{Phys. Lett. {\bf B#1}, #3 (19#2)}
\def\PRD#1#2#3{Phys. Rev. {\bf D#1}, #3 (19#2)}

\def\JMP#1#2#3{J. Math. Phys {\bf #1}, #3 (19#2)}
\def\CMP#1#2#3{Commun. Math. Phys.{\bf #1}, #3 (19#2)}
\def\RMP#1#2#3{Rev. Mod. Phys. {\bf #1}, #3 (19#2)}
\def\GRG#1#2#3{Gen. Rel. and Grav. {\bf #1}, #3 (19#2)}

\begin{document}
\begin{titlepage}
\thispagestyle{empty}
\begin{flushright}
BU-HEP-98-7 \\
DOE/ER/40561-358-INT98-00-6 \\
UW/PT-97/24 \\
hep-th/9803132 \\
\end{flushright}
\vspace{1.0cm}
\begin{center}
{\Large\bf Effective Field Theory, Black Holes, \\
and the Cosmological Constant\\
}
~\\
{Andrew G. Cohen$^a$, David B. Kaplan$^b$ and  Ann E. Nelson$^c$} \\
~\\
\noindent
{\it\ignorespaces
       (a) Department of Physics, Boston University,
           Boston, MA 02215, USA        \\
       (b) Institute for Nuclear Theory, 1550,
           University of Washington, Seattle, WA 98195-1550, USA   \\
       (c) Department of Physics 1560, University of Washington,
           Seattle, WA 98195-1560, USA\\

}
\end{center}
\vspace*{\fill}
\begin{abstract}
 
  Bekenstein has proposed the bound $S\le \pi M_P^2 L^2$ on the total
  entropy $S$ in a volume $L^3$.  This non-extensive scaling suggests
  that quantum field theory breaks down in large volume.
 To reconcile this breakdown with the success
of local quantum field theory in describing  observed particle
  phenomenology, 
we propose a relationship between UV and IR cutoffs such
that an effective field theory should be a good description of
Nature. We discuss implications for the cosmological constant problem.
We find a limitation on the accuracy which can be achieved by
conventional effective field theory: for example, the minimal
correction to $(g-2)$ for the electron from the constrained IR and UV
cutoffs is larger than the contribution from the top quark.

\end{abstract}
\vspace*{\fill}

\noindent
\parbox{0.45\textwidth}{\hrule\hfill} \\
\begin{tabular}{cl} e-mail: &  {\tt cohen@andy.bu.edu}\\
                            & {\tt dbkaplan@phys.washington.edu} \\
                            &  {\tt anelson@phys.washington.edu}
\end{tabular}

\end{titlepage}

It is generally assumed that  particle  physics  can be
accurately described by an effective field theory with  
  an ultraviolet (UV) cutoff 
less than the Planck mass $M_P$, provided that all momenta and  
field strengths are small compared with
this cutoff to the appropriate power. 
Computations performed with such effective field theories, for example
the Standard Model, have been extraordinarily successful at describing 
properties of elementary particles.
Nevertheless, considerations involving black holes suggest that the
underlying theory of Nature is not a local quantum field theory. In 
this Letter we attempt to reconcile this conclusion with the 
success of effective quantum field theory by determining the range of validity
for a local effective  field theory to be an accurate  description of
the world.
 We accomplish this by imposing a relationship between UV and infrared (IR)
cutoffs.
We will argue that this proposed IR bound does not conflict with any current
experimental success of quantum field theory, but  explains why  conventional
effective field theory estimates of the cosmological constant fail so 
miserably. 

For an effective quantum field theory in a box of size $L$ with UV
cutoff $\Lambda$
the entropy $S$ scales extensively, $S \sim L^3
\Lambda^3$\footnote{For 
example, a free Weyl
  fermion on a  lattice of size $L$ and spacing $1/\Lambda$ has
  $4^{(L\Lambda)^3}$ states and entropy $S=(L\Lambda)^3\ln 4$ 
    (ignoring lattice doublers); a
  lattice theory of bosons represented by a compact field likewise has
  entropy scaling as $(L\Lambda)^3$.}. 
However the peculiar thermodynamics of black
holes~\cite{Bekenstein,Hawking} has led Bekenstein~\cite{Bekenstein}
to postulate  
that the maximum entropy
in a box of volume $L^3$ behaves non-extensively, growing only as the
area of the box. For any $\Lambda$, there is a sufficiently large
volume for which the
entropy of an effective field theory  will exceed
the Bekenstein limit.  't Hooft~\cite{tHooft} and Susskind~\cite{Susskind}
have stressed that this result implies conventional $3+1$ dimensional
field theories vastly over-count degrees of freedom: as these field
theories are described in terms of a Lagrange density, they
have extensivity of the entropy built in.
The Bekenstein entropy bound may be satisfied in an effective field theory
if we limit the volume of the system according to
\begin{equation}
  \label{bekbound}
  L^3 \Lambda^3 \ltap S_{BH} \equiv \pi L^2 M_P^2
\end{equation}
where $S_{BH}$ is the entropy of a black hole of radius
$L$~\cite{Hawking,Bekenstein}.
Consequently the length $L$, which acts as an IR cutoff, cannot be chosen
independently of the UV cutoff, and scales as $\Lambda^{-3}$.

As startling as the Bekenstein-motivated  constraint  eq.~(\ref{bekbound}) 
seems,  there is evidence that conventional quantum field
theory  fails at an entropy well below this bound.
't Hooft has stressed that ordinary field theories
should fail on large scales if near the horizon of a
black hole~\cite{tHooft}. In the presence of even a very large
  black hole, a low energy description of particle physics is expected
  to be inadequate, since infalling particles experience Planck scale interactions
  with outgoing Hawking radiation near the horizon. 
  Furthermore, it  has been shown in string theory
that local observables  do not
necessarily commute at space-like separation  in the presence of a
black hole~\cite{LPSTU}. These problems arise even in the absence of
any large field strengths or momenta.
Local quantum field theory appears unlikely to be a good effective low energy
description of any system containing a black hole,
and should probably not attempt to describe particle states whose volume is
smaller than their corresponding Schwarzschild radius. 

An effective field theory that can saturate eq.~(\ref{bekbound})
necessarily includes many states with Schwarzschild radius much larger than
the box size. To see this,
note that a conventional effective quantum
field theory  is expected to be capable of
describing a system at a temperature $T$, provided that 
$T\le \Lambda$; so long as $T\gg 1/L$, 
such a system has thermal energy $M\sim L^3 T^4$ and
entropy $S\sim L^3 T^3$.  When  eq.~(\ref{bekbound})
is saturated, at $T\sim (M_P^2/L)^{1/3}$, the corresponding
Schwarzschild radius $L_S$ for this system is $L_{S} \sim L (L
M_P)^{2/3}\gg L$.

To avoid these difficulties  we propose
an even  stronger constraint on the 
IR cutoff $1/L$ which excludes all states that lie within their Schwarzschild 
radius.  Since the maximum energy density in the effective theory is
$\Lambda^4$, the constraint on $L$ is
\begin{equation}
\label{blueberunium}
L^3 \Lambda^4 \ltap L M_P^2\ .
\end{equation}
Here the IR cutoff scales like $\Lambda^{-2}$. This bound is far more
restrictive than eq.~(\ref{bekbound}):
when eq.~(\ref{blueberunium}) is near saturation, the entropy is
\begin{equation}
\label{oursbound}
S_{max} \simeq S_{BH}^{3/4}\ .
\end{equation}
We propose
that an effective local quantum field theory will be a good approximate 
description of physics  when
eq.~(\ref{blueberunium}) is  satisfied. This bound is more
  restrictive than eq.~(\ref{bekbound}) because we are explicitly considering
  only those states that can be described by conventional quantum
  field theory\footnote{The fact
that systems which do not contain
black holes have maximum entropy of order $S_{BH}^{3/4}$ is well known
\cite{Bekenstein,tHooft,threequarters}. The entropy of  black holes has
been explicitly counted in string theory \cite{StromingerVafa} and
M-theory
\cite{MSW} and appears to involve many states which are not describable
within ordinary field theory.}.

Can such a dramatic depletion of quantum states  
be relevant to the cosmological constant problem\footnote{Banks has also 
  argued that a drastically reduced
  number of fundamental degrees of freedom may
  be part of an explanation of the small size of the
  cosmological constant~\cite{Banks}. His explanation differs from ours
  as he uses the weaker Bekenstein bound on the UV cutoff, and assumes
  both an IR cutoff which is much larger than the present horizon, as
  well as substantial cancellations of the zero-point energies of the
  fundamental degrees of freedom due to supersymmetry in a 2+1
  dimensional ``holographic'' description.
  Horava~\cite{horava} has  proposed a model with nonextensive
  fundamental degrees  
  of freedom which also gives suppression of the cosmological constant.}? 

If the Standard Model is valid in an arbitrarily large volume up to at
least LEP energies, then the quantum
contribution to the vacuum energy density computed in perturbation theory
is $\sim (100~\GeV)^4$. The empirical bound on the
cosmological constant corresponds to a vacuum energy density $\ltap
(10^{-2.5}~\eV)^4$. Conventionally this discrepancy
is explained by either unknown
physics at high energies which
conspires to cancel this vacuum contribution
to enormous precision, or else new physics at $\sim
10^{-2.5}~\eV$ which adjusts to cancel the vacuum energy while being
devious enough to escape detection~\cite{Weinberg,newphyssol}.

There is however a third possibility---that the usual perturbative
computation of the quantum correction to the 
vacuum energy density, which assumes no infrared limitation to the quantum 
field theory, is incorrect. There is in fact no evidence that fields
at 
present experimental  energies
can  fluctuate independently over a region as large as our horizon. 
In fact, if we
choose an IR cutoff comparable to the current horizon size, the
corresponding UV cutoff from eq.~(\ref{blueberunium}) is $\Lambda \sim 10^{-2.5}~\eV$ and the resulting
quantum energy density of $\Lambda^4$ requires no cancellation to be 
consistent with current
bounds. This observation does not predict the 
cosmological constant's value,  as one can always add a constant to the 
quantum contribution.  However it does eliminate the need for fine-tuning.


The peculiar relationship between IR and UV cutoffs
eq.~(\ref{blueberunium}) is, in principle,
testable as it limits
the successful application of quantum field theory to
experiment. For instance if we wish to search for new physics (coming
from new interactions or particles at high energies which do not
violate low energy symmetries)
using high precision experiments at low energies $p$, there is a
maximal energy scale that can be probed without incorporating effects
beyond conventional quantum field theory. Surprisingly, this scale
depends on $p$, and can be much lower than $M_P$.

In order to perform an effective field theory calculation we 
simultaneously impose a UV and an IR cutoff consistent with
eq.~(\ref{blueberunium}). There will be small discrepancies between
such a calculation and a conventional one performed in
an infinite box. 
Such a discrepancy  can be of
interest when trying to discover new physics through radiative corrections.
For example, consider $(g-2)$ for the electron.
The UV and IR cutoffs that we must impose 
each lead to corrections to the usual calculation,
whose total size is
\begin{equation}
\label{deltag}
\delta(g-2)\sim {\alpha\over \pi}
\left[\left({m_e\over \Lambda}\right)^2 +\left({1\over m_e L}\right)^2\right]
\end{equation}
If we were able to choose $L$ independently of $\Lambda$ we would
simply ignore the IR corrections. However we must now comply with
eq.~(\ref{blueberunium}). Substituting this constraint on $L$ gives
\begin{equation}
\label{deltag2}
\delta(g-2) \gtap {\alpha\over \pi}
\left[\left({m_e\over \Lambda}\right)^2 +\left({\Lambda^2\over m_e
M_P}\right)^2\right]\ .
\end{equation}
This uncertainty in our calculation is minimized by choosing the
UV cutoff to be $\Lambda \sim ( m_e^2 M_P)^{1/3}\sim 14$~TeV, so that
\begin{equation}
\label{deltagmin}
\delta_{min}(g-2)\sim {\alpha\over \pi}\left( {m_e\over M_P}\right)^{2/3}\sim
{\alpha\over \pi}\times 10^{-15} \ .
\end{equation}
While still  small, this deviation is
far larger than the usual effects one would ascribe to gravity.  In
fact, the minimal discrepancy in the
calculation of $(g-2)$ that arises in this way is equivalent to the
contribution
from a lepton of mass $M\sim 100$~GeV, and is roughly twice
 the contribution to  $(g-2)$ from the top quark\footnote{The conventional
contribution of a new heavy lepton of mass $M$ to $(g-2)$ is \cite{combley}
$\Delta(g-2)={(1/45)}\left({\alpha/ \pi}\right)^2 \left({m_e/ M}\right)^2$ ,
where $m_e$ is the electron mass.}.
These effects are enormously larger than conventional estimates of Planck
scale corrections which are of order $(m_e/M_P)^2\sim 10^{-44}$.

More generally, we may consider
processes of characteristic energy $p$ which receive
contributions from dimension $D$ operators  with $D>4$,
characterizing new physics.
The correction due to a finite UV cutoff is of order
$(\alpha/\pi)\ (p/\Lambda)^{(D-4)}$.
The required IR cutoff $L\ltap  M_P/\Lambda^2$ leads
to additional corrections $\sim (\alpha/\pi)\
(1/(L^2 p^2))$, which are at least as big as
$(\alpha/\pi)\ (\Lambda^2/(p M_P))^2$, according to
our constraint eq.~(\ref{blueberunium}).
Minimization
of this theoretical uncertainty
occurs for a UV cutoff $\Lambda\sim p( M_P/p )^{2/D}$.
Thus in a given experiment there is a maximum
energy scale that can be probed and a maximum accuracy that can be
achieved using conventional quantum field theory, with the energy scale
depending on $M_P$ to a remarkably small fractional power.
For operators of dimension 5 this scale is  $M_P^{2/5}$, while for
operators of dimension six it is $M_P^{1/3}$.

Note that the relative size of these effects grows with $p$.
When $p$ is the weak scale and the effective theory is
the Standard Model, new physics at short distances appears in the
effective theory through
dimension six operators, and the maximum energy scale that can be
conventionally probed is $10^8~\GeV$, with a corresponding uncertainty of
$10^{-13}$. If the new high energy physics appears through dimension five
operators  the maximum energy
scale would be $10^9~\GeV$ with an uncertainty of $10^{-9}$.

We might worry that the low scale which can be probed by electroweak
physics eliminates the possibility of computing
coupling constant unification,
which involves an energy scale $M_{GUT} \sim 10^{16}~\GeV$.
However it is still possible to consider running of dimension four
operators up to energies as high as $M_P$.
In
order to compute coupling constant running in the presence of our IR
and UV cutoffs, we may use a renormalization group treatment, matching
the S-matrices of two theories with parameters $\{L, \Lambda\}$ and
$\{L',\Lambda'\}$ (each of which obey eq.~(\ref{blueberunium})) in their
combined domain of validity. There is an inherent uncertainty
in the beta function at a given energy scale due to the effects of these
cutoffs. Choosing $\{L, \Lambda\}$ at each energy scale to minimize
the uncertainty\footnote{In the Standard Model or in the MSSM, UV cutoff
  effects enter via dimension-six
  operators involving gauge field strengths, such 
  as $G^a_{\mu\nu}\partial^2 G^{a\mu\nu}$.}, 
leads  to corrections of the relation
between the unified coupling at $M_{GUT}$ and the Standard Model gauge
couplings at $M_Z$:
\begin{equation}
  {4\pi\over \alpha_i(M_Z)} = {4\pi\over \alpha_{GUT}(M_{GUT})} +
    b_i\ln{m_z\over M_{GUT}} + {\cal O}((M_{GUT}/M_P)^{2/3})\ .
\end{equation}
These corrections are small, but comparable to  the usual 2-loop
 corrections, and are not obviously out of experimental reach. 
Thus if one had a compelling reason to believe in a particular 
GUT with a unification scale well below $M_P$, one might be able to  use 
visible deviations from its low energy coupling constant predictions
as evidence for  the limitations of quantum field theory proposed here. 

As our renormalization group analysis of gauge coupling
  flow differs from the conventional analysis by only small corrections, one
  might expect to obtain conventional results for the RG
  flow of the vacuum energy as well,  recovering the usual
  fine-tuning problem associated with the cosmological constant, arising from
  quartic divergences. 
  However, in order to match two theories with cutoffs  $\{L, \Lambda\}$ and
$\{L',\Lambda'\}$ by requiring that they reproduce the same physical
vacuum energy density $\lambda$ --- by comparing graviton propagators
about a flat metric, for example --- the lengths $L$ and $L'$ both must be
larger than the length scale $M_p /\sqrt{\lambda}$, in order to avoid
spurious finite volume effects.  This implies that one cannot perform
the RG scaling to UV cutoffs larger $\lambda^{(1/4)}$, and that
consequently one never sees a fine tuning problem for the vacuum
energy.

It is conceivable that black holes and their interactions with
  particles can be described by some effective field 
theory, eliminating the motivation for the bound of eq.~(\ref{blueberunium}). 
It remains difficult to understand the necessary non-extensive
behavior of the entropy  
without {\em some} infrared limitation of effective field theory
at least as strong as the 
Bekenstein-motivated bound of
eq.~(\ref{bekbound}). 
However  even  this latter bound leads to conclusions qualitatively similar
to those above.
For example experiments at a scale $p$ sensitive to new
physics which arises through dimension $D$
operators ($D>4$) would be limited to probing energies below $\Lambda\sim
\left(p^{D-2}M_P^4\right)^{1/(D+2)}$, and the
maximum theoretical
accuracy would be $\sim(\alpha/\pi)(p/M_P)^{4(D-4)/(D+2)}$.

Both bounds give relatively large corrections to effective field
theory computations compared to conventional computable quantum
gravitational effects.  The latter are generally expected to be
suppressed by integral powers of $M_P$; such expectations are born
out by explicit constructions of effective field theories from string
theory \cite{string}. 

It is tempting to consider a less drastic solution: patching up
  conventional effective field theory (with 
  a Planck scale UV  cutoff and no IR cutoff) by eliminating ``by
  hand'' 
those
states corresponding to black holes.  
We do not know how to  prune a Hilbert space in
this manner; the result would likely be a bizarre,
nonlocal theory. Still,  one could
  imagine that even though most 
of the degrees of freedom in an effective field theory in an
arbitrarily large box have no sensible physical interpretation, for
some reason the theory accurately describes the properties of few
particle states. This would leave conventional calculations which
contain no intermediate states approaching black hole
  formation unchanged
to low orders in perturbation theory,  while rejecting the numerous states predicted by the 
  same theory which lie within their own Schwarzschild radius,
however there would be drastic effects on thermal distributions even at
temperatures $T \ll \Lambda$.
 Instead, our main assumption is that a local effective field
theory which 
correctly describes all single particle states with  momenta up to
$p\sim \Lambda$, should also describe multi-particle excitations, and
would have a normal density matrix for
thermal distributions with 
$T\ll\Lambda$.   While  conventional, this assumption may not be
valid when the underlying theory is not local. The alternative
that an effective field theory can be valid up to a scale $\Lambda$
for certain calculations, but fail to correctly describe a thermal system at
temperature $T \ll \Lambda$, seems at least as strange as our
assumption.

In conclusion, many different results about the physics of black holes
imply that, in the presence of quantum gravity, there are no
fundamental extensive degrees of freedom. Furthermore, considerations
of the maximum possible entropy of systems which do not contain black
holes suggest that ordinary quantum field theory may not be valid for
arbitrarily large volumes, but would apply provided the UV and IR
cutoffs satisfy a bound given by eq.~(\ref{blueberunium}).  The
experimental success of quantum field theory survives,
as long as this effective theory is not applied to calculations which
simultaneously require both a low infrared cutoff and an overly high
UV cutoff. The simultaneous UV and IR sensitivity of computations
relevant for current laboratory experiments never comes close to
requiring cutoffs which violate eq.~(\ref{blueberunium}).  In
contrast, the computation of the quantum contribution to the vacuum
energy of the visible universe within quantum field theory requires a
UV cutoff of less then $10^{-2.5}$~eV. With this cutoff, 
no fine-tuned cancellation of the cosmological constant is
required. Recognition that quantum 
  field theory vastly overcounts states can help resolve the enormous 
discrepancy between
conventional estimates of the vacuum energy and the observed cosmological
constant and eliminate a celebrated fine-tuning problem.

\bigskip\bigskip\noindent
{\bf Acknowledgements.}
We gratefully acknowledge Tom Banks for very useful discussions, and the
Aspen Center for Physics, where this work was initiated. We also thankfully
acknowledge critical correspondence  from Finn Larsen, Steve Giddings,
Petr Horava, 
Juan Maldacena, Yossi Nir, Joe Polchinski and an anonymous referee. 
A.G.C. is supported in part by DOE grant
\#DE-FG02-91ER40676; D.B.K. is supported in part by DOE grant
\#DOE-ER-40561;
A.E.N. is supported in part by DOE grant \#DE-FG03-96ER40956.


\newpage
\begin{thebibliography}{9}

\bibitem{Bekenstein} 
J. D. Bekenstein,
{\it Black Holes and Entropy}, \PRD{7}{73}{2333};
{\it Generalized Second Law of Thermodynamics in Black
Hole Physics}, \PRD{9}{74}{3292};
{\it A Universal Upper Bound On The Entropy to Energy
Ratio for Bounded Systems}, \PRD{23}{81}{287};
{\it Entropy Bounds and Black
Hole Remnants}, gr-qc/9307035, \PRD{49}{94}{1912}.

\bibitem{Hawking}
S. W. Hawking, {\it Particle Creation by Black Holes},
\CMP{43}{75}{199};
{\it Black Holes and Thermodynamics}, \PRD{13}{76}{191}.

\bibitem{tHooft}
G. 't Hooft, {\it Dimensional Reduction in Quantum
Gravity}, gr-qc/9310026, published in {\bf Salamfestschrift: a
collection of talks},  Eds.  A.   Ali, J. Ellis, and S. Randjbar-Daemi,
 World Scientific, 1993.

\bibitem{Susskind}
L. Susskind, {\it The World as a Hologram}, hep-th/9409089,
\JMP{36}{94}{6377}.

\bibitem{LPSTU}
D. A. Lowe, J. Polchinski, L. Susskind, L. Thorlacius, and J. Uglum,
{\it Black Hole Complementarity Versus Locality}, hep-th/9506138,
\PRD{52}{95}{6997}.

\bibitem{threequarters}
D. N. Page, Physics Today {\bf 30}, 1 (1977), {\it Black Hole Formation
in a Box}, \GRG{13}{81}{1117};
R. D. Sorkin, R. M. Wald, and Z. Z. Jiu, {\it Entropy of Self
Gravitating Radiation}, \GRG{13}{81}{1127}.

\bibitem{StromingerVafa}
A. Strominger and C. Vafa, {\it Microscopic Origin of the Bekenstein-Hawking
Entropy}, hep-th/9601029, \PLB{379}{96}{99}.

\bibitem{MSW} J. Maldacena, A. Strominger, and E. Witten, {\it Black Hole
    Entropy in M-Theory}, hep-th/9711053.

\bibitem{Banks}
T. Banks, {\it SUSY Breaking, Cosmology, Vacuum Selection and the
Cosmological Constant in String Theory}, hep-th/9601151.

\bibitem{horava}
P. Horava, {\it M Theory as a Holographic Field Theory},
  hep-th/9712130. 

\bibitem{Weinberg}
S. Weinberg, {\it The Cosmological Constant Problem}, \RMP{61}{89}{1},
and references therein.

\bibitem{newphyssol}
R. Sundrum, {\it Towards an Effective Particle String Resolution of
the Cosmological Constant Problem}, hep-ph/9708329.

\bibitem{combley} For example, F. H. Combley, {\it $(g-2)$ Factors for
Muon and Electron and the Consequences for QED}, Rep. Prog. Phys.
{\bf 42} (1979) 1889.

\bibitem{string}
J. Maldacena, private communication.

\end{thebibliography}
\end{document}